\documentclass[aps,prb,twocolumn,floatfix,showpacs]{revtex4}
\usepackage{epsf}
\usepackage{graphicx}           
\usepackage{amssymb}
\usepackage{amsmath}

\begin{document}

\title{On the dimerized phase in the cross-coupled antiferromagnetic spin ladder}

\author{G. Barcza$^{1,2}$, \"O.~Legeza$^{1,2}$, R.~M.~Noack$^1$, and J. S\'olyom$^2$} 
\affiliation{$^1$ Fachbereich Physik,
  Philipps-Universit\"at Marburg, 35032 Marburg, Germany\\
  $^2$ Research Institute for Solid State Physics, H-1525 Budapest, P.\ O.\ Box 49, Hungary}

\date{\today}

\vskip -8pt

\begin{abstract}
We revisit the phase diagram of the frustrated $s$=1/2 spin ladder
with antiferromagnetic rung and diagonal couplings. 
In particular, we reexamine the  
evidence for the columnar dimer phase, which has been predicted from
analytic treatment of the model and has been claimed to be found in numerical
calculations. 
By considering longer chains and by keeping more states than in
previous work using the density-matrix renormalization group,  
we show that the 
numerical evidence presented previously 
for the existence of the dimerized phase 
is not unambiguous 
in view of the present more careful analysis. 
While we cannot completely rule out the possibility of a dimerized
phase in the cross-coupled ladder, we do set limits on the maximum
possible value of the dimer order parameter that are much smaller than
those found previously.
\end{abstract}


\maketitle

\section{Introduction}

In spin chains,
frustration due to competing antiferromagnetic nearest- and comparably
strong  next-nearest-neighbor coupling 
leads to the appearance of a 
new gapped, dimerized phase\cite{hald,affleck} whose properties are markedly
different from that of the uniform spin-liquid 
state of the antiferromagnetic 
spin-1/2 chain. A much richer phase diagram is found in spin ladders, where 
the couplings between nearest-neighbor spins situated along the legs and on 
the rungs of the ladder may compete. These phases and the quantum phase 
transitions between them have been extensively studied recently, both
experimentally and theoretically.\cite{frustrated} 
The theoretical approaches include both 
analytic and numerical work. The analytic treatment of the bosonized form of 
the spin Hamiltonian has been widely used to predict the phase diagram. 
\cite{cg,shelton,nt,chp,og,weihong,nge,gt,aen,gene,oleg}  
There is general agreement that, in two-leg spin ladders, the most relevant 
operator generates either a topologically ordered rung-singlet or a
Haldane phase,
depending on the relative strength of the intraleg and interleg couplings. 

The competition between the couplings is, however, a delicate problem when 
the intraleg and the interleg couplings are of the same order of magnitude, 
and the type of order 
is, in some cases, determined by marginally relevant terms generated 
within 
the renormalization group (RG) procedure. Among the many 
terms generated, 
operators leading to 
dimerization appear. 
On this basis,\cite{oleg} it was proposed that a dimerized phase
appears between
the topologically ordered rung-singlet and the Haldane phases for the so-called 
cross-coupled ladder, in which spins are coupled not only along the legs and 
across the rungs but 
also diagonally, i.e., on opposite legs and on different, neighboring rungs.

The existence of an analogous 
dimerized phase induced by 
four-spin exchange terms has been proposed some time ago.\cite{nt} 
This dimerized phase has indeed been found to exist 
numerically.\cite{lfs,kolezhuk,pati,muller,lauchli,momoi} 
It is also well-established 
that a dimerized phase exists when a second-neighbor interaction on the
legs of the 
ladder are present 
\cite{honecker} or when the interleg coupling is 
ferromagnetic.\cite{hikihara} The existence of the dimerized phase has, 
however, been 
disputed for ladders with pairwise spin 
interactions only,\cite{wang,hung,kim2008} while other work has presented
evidence supporting the presence of this phase.\cite{liu, hikihara}
 
The density-matrix renormalization group (DMRG) procedure invented by 
White\cite{white} is a particularly useful numerical method to study such
problems because it makes possible the 
high-precision calculation of the ground-state 
properties and low-lying excitations of low-dimensional quantum
systems. 
The DMRG 
has indeed been widely used \cite{lfs,fath,hf,hhr,fouet,meyer,sheng} to explore 
the possible phases of spin ladders. 
The length of the ladders from which 
finite-size scaling is done and the accuracy of the DMRG calculation
can be, however, decisive.
Motivated mainly by the recent work of Hikihara and 
Starykh\cite{hikihara}, we have undertaken a careful analysis of the 
numerical 
{data to see if there is indeed compelling} 
evidence for the dimerized phase in the cross-coupled ladder
using the DMRG procedure.

The paper is organized as follows. In Sec.~II, we summarize earlier results 
on the phase diagram of the cross-coupled frustrated antiferromagnetic
spin-1/2 ladder. The aspects of the DMRG procedure that determine
its accuracy are briefly discussed in Sec.~III. Our numerical results are
presented in Sec.~IV. Finally Sec.~V contains our conclusions.

\section{Previous results on the cross-coupled spin ladder}

\begin{figure}[htb]
\includegraphics[trim = 0cm 18.5cm 0cm 0cm, clip, width=8.1cm]{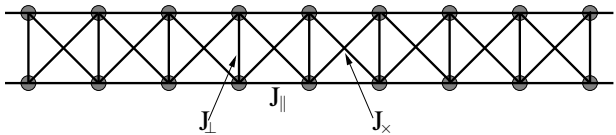}
\caption{(Color online) Depiction 
of the cross-coupled ladder. } 
\label{fig:composite}
\end{figure}

We treat a two-leg spin-1/2 ladder with pairwise Heisenberg couplings
along the legs, across the rungs, and diagonally between sites on
adjacent rungs, but on opposite legs.
The Hamiltonian can be written
\begin{equation}
    {\mathcal H} = \sum_{i} {\mathcal H}_{i,i+1}\,,
\label{eq:ham}
\end{equation}
where
\begin{eqnarray}
   {\mathcal H}_{i,i+1} &=& J_{\|} 
   \left( \boldsymbol{S}_{1,i} \cdot \boldsymbol{S}_{1,i+1}  + 
  \boldsymbol{S}_{2,i} \cdot \boldsymbol{S}_{2,i+1}\right)  \nonumber \\
& & + \textstyle{\frac{1}{2}}J_{\bot} 
   \left( \boldsymbol{S}_{1,i} \cdot \boldsymbol{S}_{2,i} +  
  \boldsymbol{S}_{1,i+1} \cdot \boldsymbol{S}_{2,i+1}\right)  \\
&& + J_{\times} \left( \boldsymbol{S}_{1,i} \cdot
\boldsymbol{S}_{2,i+1} + \boldsymbol{S}_{2,i} \cdot 
\boldsymbol{S}_{1,i+1} \right) \,,   \nonumber 
\end{eqnarray}
and $\boldsymbol{S}_{1,i}$ and $\boldsymbol{S}_{2,i}$ are the spin 
operators on rung $i$ and legs 1 and 2, respectively. 
A depiction of the couplings of the cross-coupled ladder is 
shown in Fig.~\ref{fig:composite}.
Here we will consider only 
antiferromagnetic coupling and set the intraleg coupling  
to unity, $J_{\|}=1$ which we take as 
the energy scale in the problem.
For $J_{\times}=0$, it has been 
established that the entire half-line $0 < J_{\bot} < \infty$ of the
ground-state phase diagram, 
is continuously connected,\cite{wns,shelton} i.e., no phase
transition occurs for any finite value of $J_{\bot}$. 
Since the
configuration that dominates the 
ground-state wave function is the product of rung singlets,
this phase is often referred to as the {\sl rung-singlet}
(RS) phase. 
Conversely, for $J_{\bot}=0$,
the entire half-line $0 < J_{\times} < \infty$ is continuously
connected to the {\sl Haldane phase} \cite{haldane} of the spin-1 chain and
the system is known to have a gapped spectrum of spin-1 magnons. \cite{lfs,ls}  

When both $J_{\bot}$ and $J_{\times}$ are finite, frustration due to 
competition between the intraleg and the rung or diagonal interleg couplings 
potentially leads to the formation of new phases. 
Indications for such new phases are given by analytic calculations
which can be performed 
for weak interchain couplings using the bosonized form of the
Hamiltonian. 
The leading term in the Hamiltonian is the coupling between the
staggered magnetizations of the legs:
\begin{equation}
    (J_{\bot}-2 J_{\times})\boldsymbol{n}_1 \cdot \boldsymbol{n}_2  \,.
\end{equation}
This term generates a gap in the spectrum irrespective of the sign of
the coefficient $J_{\bot}-2 J_{\times}$. 
Rung singlets are generated for $J_{\bot}-2 J_{\times} > 0$,
while the Haldane phase appears for $J_{\bot}-2 J_{\times} < 0$. 
The first numerical results on the model \cite{weihong,fath} were
consistent with a first-order transition for both weak and strong
interchain couplings. 
Subsequent work 
has suggested that the transition is actually continuous when the interchain 
coupling is weak, becoming first-order only for stronger interchain 
coupling.\cite{wang,hung} 
This scenario has been supported further  
by an analysis based on quantum information entropies.\cite{kim2008} 
Although two critical values of the coupling were found in the
two-site entropy function for
finite systems in this work, it was shown that they
scale to a single value in the thermodynamic limit. 
It was therefore argued that a direct transition between
the rung singlet and the Haldane phases takes place. 

When $J_{\bot} = 2J_{\times}$, 
the most relevant term is suppressed at the transition between the two
gapped  phases and other, less relevant terms may play a role. 
In particular, such terms 
lead to the prediction 
of a spontaneously dimerized
phase.\cite{oleg,kim2008,hikihara}
The boundaries of this dimerized phase have been estimated to lie at
\begin{equation}
    J_{\bot} = 2 J_{\times} - \frac{5}{\pi^2} J^2_{\times}  \quad \mbox{and}
	   \quad    J_{\bot} = 2 J_{\times} - \frac{1}{\pi^2} J^2_{\times}  \,.
\end{equation}
The most recent numerical results, however, indicate a
much narrower range for the existence of the dimerized
phase.\cite{liu,hikihara}  
In Ref.~\onlinecite{liu} the dimerized phase was 
claimed to exist in the range  
$0.373 \leq J_{\bot} \leq 0.386$ for $J_{\times} = 0.2$. 
{ The authors of Ref.~\onlinecite{hikihara} chose a single point in the
middle of this range, namely
$J_{\bot} = 0.38$, $J_{\times} = 0.2$ to demonstrate the existence of the
dimerized phase. In what follows, 
we will also restrict ourselves to the vicinity of this single point
in parameter 
space and will carry out a careful numerical analysis of the dimer
order parameter in the range $0.36\leq J_{\bot}\leq 0.4$ for $J_{\times} = 0.2$. 
}
\section{Numerical procedure}

\subsection{Factors determining the accuracy of the DMRG procedure}

In the standard DMRG procedure, the chain is divided into two blocks
and two extra sites.
As the two sites are incorporated into larger blocks during the steps of
the renormalization procedure, only a fraction of the new block states
are kept; the others are discarded. 
Usually the number of block states, $M$, kept in the truncation
procedure ranges from a few hundred to a few thousand
and is set  
according to the accuracy required and the available computational
resources.\cite{schollwock,manmana,hallberg}

As we have pointed out in previous work, keeping
the number of block states constant for increasingly large system
sizes leads to increasing error, \cite{legeza02,legeza03} 
i.e., 
the scaling of block entropy with system size is not taken into
account.\cite{vidal03} 
In a consistent calculation, the number of block states must be
increased as system size is increased. 
The finite-size scaling to infinite system size must  
be accompained by a systematic increase of $M$, and the 
$M \rightarrow \infty$ limit must be taken. 
We have shown that the proper choice of $M$ can be 
determined by the so-called Dynamic Block State Selection (DBSS) 
approach,\cite{legeza02,legeza03} in which the threshold value of the quantum 
information loss $\chi$ is fixed {\em a priori}. 
In this approach, the number of block 
states to be kept is obtained from the criterion $s-s^{\rm Trunc}\leq\chi$, where
$s$ and $s^{\rm Trunc}$ stand for the block entropy before and after the 
truncation, respectively. The minimum number of block states $M_{\rm min}$ is 
set prior to the calculation, and the maximum number of block states
needed to achieve the desired 
accuracy, $M_{\rm max}$, is monitored
during the DMRG process. 
This allows for a rigorous control of the numerical accuracy and gives 
a stable extrapolation to $\chi = 0$.

\subsection{Choice of a proper order parameter: advantages and limitations}

In an earlier work\cite{gene} we used the generalization of the 
so-called hidden topological order,\cite{dennijs} $O$, to two-leg
ladders,
\begin{eqnarray}
     {\mathcal O}_{\text{odd}} & = & - \lim_{|i-j| \rightarrow \infty }
	 \Bigg\langle  (S_{i,1}^z + S_{i,2}^z) \\
  &&  \times \exp \left[ i \pi \sum_{l = i+1}^{j-1}
	 (S_{l,1}^z + S_{l,2}^z) \right] (S_{j,1}^z + S_{j,2}^z) \Bigg\rangle \,, 
	 \nonumber
\end{eqnarray}
and
\begin{eqnarray}
     {\mathcal O}_{\text{even}} & = & - \lim_{|i-j| \rightarrow \infty }
	 \Bigg\langle  (S_{i+1,1}^z + S_{i,2}^z) \\
  &&  \times \exp \left[ i \pi \sum_{l = i+1}^{j-1}
	 (S_{l+1,1}^z + S_{l,2}^z) \right] (S_{j+1,1}^z + S_{j,2}^z) \Bigg\rangle \,, 
\nonumber
\end{eqnarray}
to identify various spin-liquid phases.\cite{fath} The advantage of these 
specially constructed order parameters is that only one of them can be finite in 
a given phase and they both vanish for critical systems. Unfortunately, the 
rung-singlet and dimerized phases would fall into the same topological sector 
because $O_{\rm even}$ is finite for both cases. Thus, they
cannot be distinguished by the topological order. 

A similar problem arises for the so-called $z$ operators,\cite{nakamura01,nakamura02} 
which can also be used to distinguish different odd- or even-parity valence-bond-solid 
(VBS) states. The operator
\begin{equation}
     z_{\text{rung}} = \left\langle \exp \left[ i \frac{2\pi}{N} \sum_{l = 1}^{N}
	 l (S_{1,l}^z + S_{2,l}^z) \right] \right\rangle\,,  
\end{equation}
defined for ladders with periodic boundary conditions, converges to $+1$ in the 
rung-singlet phase and to $-1$ in the Haldane phase, while the operator
\begin{equation}
     z_{\text{diag}} = \left\langle \exp \left[ i \frac{2\pi}{N} \sum_{l = 1}^{N}
	 l (S_{1,l+1}^z + S_{2,l}^z) \right] \right\rangle  
\end{equation}
converges to $-1$ and $+1$, respectively, in the two topologically distinct phases. 
They give a clear indication of the parity of the number of singlet bonds 
``cut" by a line between adjacent rungs, but they do not distinguish 
between the columnar-dimer phase and the rung-singlet phase.

Since we are looking for a possible columnar-dimer phase only, a more 
natural quantity to examine is the local dimer order parameter $(D)$ defined as 
\begin{equation}
D = \lim_{N \to \infty} |D(N)|
\label{eq:dimerord}
\end{equation}
with
\begin{equation}
D(N) = \langle{\cal H}_{N/2-1,N/2}\rangle - \langle{\cal H}_{N/2,N/2+1}\rangle\,.
\end{equation} 
Thus, dimerization is measured as the alternation in the bond energy
in the middle of a  sufficiently long open chain. 
Here $D$ is finite when the translational symmetry of the 
Hamiltonian is spontaneously broken in the thermodynamic limit. 
It is worth noting that our definition is slightly different from
the one used in  Ref.~\onlinecite{hikihara} for the columnar phase,
where only the intraleg part of the spin couplings was included. 
{ Although, quantitatively the two definitions give different values
for the order parameter, the qualitative
behavior is the same.}

For noncritical models with a finite correlation length, the end effects decay 
exponentially and the local quantity $D(N)$ is expected to vary to
leading order as
\begin{equation}
    D(N) = D + a N^{-\beta} \exp(-N/2\xi).
    \label{eq:scale_D}
\end{equation}
This behavior is qualitatively similar to the scaling of the gap for
periodic boundary conditions (PBC), 
except that the scaling variable is the distance of the middle of the
chain from the boundary,
$N/2$, and the exponent $\beta$ of the algebraic prefactor is, 
{\em a priori}, unknown. 
Here $\xi$ stands for the correlation length, which is finite for
gapped systems. 
The exponential convergence of local quantities such as the dimer
order parameter  $D(N)$ makes the extrapolation to the thermodynamic
limit more reliable.\cite{buchta2005} 
In cases when the available data for $D(N)$ 
not show a tendency to go to a finite value 
as a function of $1/N$, we also carry out  
extrapolations without the
exponential term  in Eq.~(\ref{eq:scale_D}). 
In addition, in order to obtain an upper bound for $D$, 
we also use a second order polynomial fit in $1/N$.   
Since the numerical accuracy is of crucial importance in the present
study, we analyze the behavior of the dimer order parameter as a function
of $M$ and $\chi$ systematically.
 
\section{Numerical results}

\begin{figure}[htb]
\includegraphics[scale=0.65]{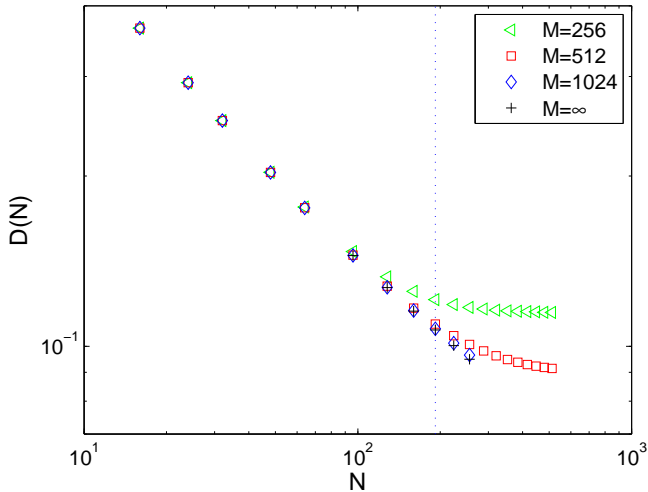}
\caption{(Color online) 
  Dimer order parameter $D(N)$ plotted as a function of $N$ on a
  log-log scale 
  for various fixed numbers of block states ($M$)
  for $J_{\bot}=0.38, J_{\times}=0.2$.  
  The dotted line indicates $N=192$, the largest system studied
  previously. 
  The $M=\infty$ extrapolation  
  is taken from the fits depicted in Fig.~\ref{fig:d-m}(a).} 
\label{fig:d-n}
\end{figure}

As mentioned above, the dimerized phase exists, if it exists at all,
in a narrow region around $J_{\bot}=0.38, J_{\times}=0.2$. 
We first calculate the length-dependent $D(N)$ using open boundary
conditions (OBC) on systems of up to $N=512$ rungs in steps of $32$.
We perform four DMRG 
sweeps with fixed numbers of block states, $M=256$ and $512$. 
Note that the longest system studied in Ref.~\onlinecite{hikihara}
had $N=192$ rungs. 
The results for the dimer order parameter at 
$J_{\bot}=0.38, J_{\times}=0.2$ 
are shown on a log-log scale in 
Fig.~\ref{fig:d-n} to allow for direct comparison with Fig.\ 10 of 
Ref.~\onlinecite{hikihara}.  
The data points calculated at fixed $M$, $M= 256$ or $M=512$,  
exhibit a clear tendency to go to a finite value as a function of $N$
in the infinite-chains limit. 
The flat region develops for $M=256$ at chain length ($N \simeq 128$)
and 
for $M=512$ when $N$ exceeds about $256$. 
This flattening of the dimer order parameter with increasing $N$ 
was  
interpreted as evidence for the presence of a dimerized phase.
 
\begin{figure}[htb]
\includegraphics[scale=0.65]{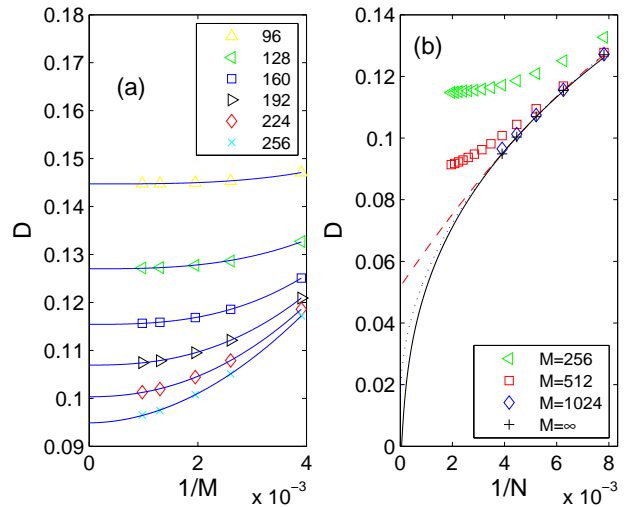}
\caption{(Color online) (a) Scaling of the dimer order parameter as a function 
of the number of block states retained, $M$, for various chain lengths.
The solid line is a fit using the function  $D(M,N) = D + a(N) M^{-\beta(N)}$. 
(b)  
The $D(N)$ data of Fig.~\ref{fig:d-n}  
plotted as a function of $1/N$. 
The solid line is a fit to the $D(M\rightarrow \infty,N)$ data 
using Eq.~\eqref{eq:scale_D}. The dotted line is a similar fit without the 
exponential factor, while the dashed line is a fit to a second-order polynomial 
in $1/N$.}
\label{fig:d-m}
\end{figure}

In order to see what happens for even larger values of $M$, we 
have carried out calculations keeping $M = 1024$ block states for
ladders of up to $N=256$ rungs. As can be seen in Fig.~\ref{fig:d-n},  
the dimer order parameter does not 
tend to a finite value with $N$ at the available system sizes. 
To carry out the standard extrapolation scheme, 
we have fitted the $M$-dependent $D(N)$ for various system sizes as a function 
of $1/M$ using the function
\begin{equation}
    D(M,N)=D(N) + a(N) M^{-\beta(N)}  \,, 
\label{eq:scale_Dm}
\end{equation}	 
where $D(N),a(N)$, and $\beta(N)$ are free fitting parameters.
These fits are shown in Fig.~\ref{fig:d-m}(a).

The finite-size scaling of $D(N)$ can be better seen if this quantity 
is plotted as a function of $1/N$ rather than  
as a function of $N$. 
Such a plot 
is shown in Fig.~\ref{fig:d-m}(b), including 
$D(M,N)$ for fixed values of $M$.
While the data points for
$M=256$ and $M=512$ behave as if they would go 
to a finite value in the $N \rightarrow \infty$
limit, the data available for $M=1024$ ($N \leq 256$) do not show an 
upward curvature in $1/N$. 
The same is true for the extrapolated values of the
dimer order parameter indicated by the solid line in Fig.~\ref{fig:d-m}(b). 
Extrapolation using Eq.~(\ref{eq:scale_D}) yields $D(M\rightarrow\infty, 
N\rightarrow\infty) = 0.003(2)$ and a very large correlation length,
$\xi=5972$. 
Since a flattening off is not apparent in the curves, we have repeated the fit  
without the exponential term, yielding 
$D(M\rightarrow\infty, N\rightarrow\infty) = 0.013(5)$. 
In order to obtain an upper bound for $D$, we have also 
carried out a fit to 
a second-order polynomial in $1/N$. 
The extrapolated value of $D_{\rm ub}(M\rightarrow\infty,
N\rightarrow\infty)$ is $0.052(2)$ 
but this fit, indicated by the dashed line,
clearly overestimates the bond order parameter significantly. 

\begin{figure}[htb]
\includegraphics[scale=0.65]{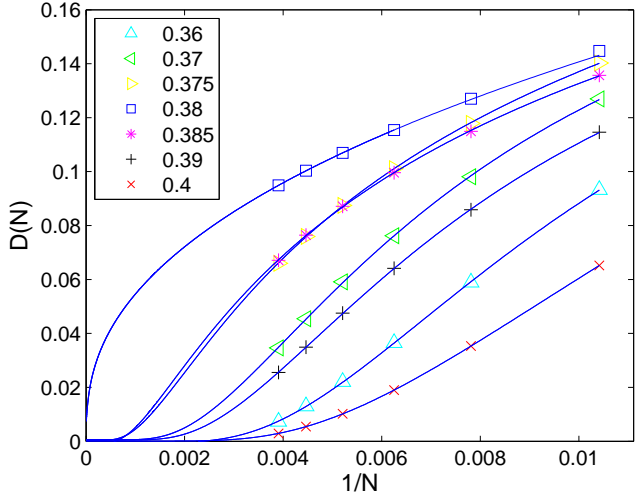}
\caption{(Color online) Scaling of $D(M\rightarrow \infty,N)$ as a function 
of $1/N$ for various $J_{\bot}$ values in the vicinity of $J_{\bot}=0.38$ keeping
$J_{\times}=0.2$ fixed. 
The solid line is a fit to the $D(M\rightarrow \infty,N)$ data 
using Eq.~\eqref{eq:scale_D}.  
}
\label{fig:d-m-n-all}
\end{figure}

{ By repeating the same analysis for several points in the vicinity of $J_{\bot}=0.38$
keeping $J_{\times}=0.2$ fixed we have also studied the finite-size scaling  
of $D(N)$ in more detail and  
determined $D(M\rightarrow\infty, 
N\rightarrow\infty)$. The results are shown in Fig.~\ref{fig:d-m-n-all}. 
For all data points 
extrapolation using Eq.~(\ref{eq:scale_D}) yields $D(M\rightarrow\infty, 
N\rightarrow\infty) \simeq \mathcal{O}(10^{-4})$ and a correlation length of the order
$10^2$, except for $J_{\bot}=0.38$. It is obvious from the figure that the inflection point shows up only at larger
system sizes as $J_{\bot}=0.38$ is approached, thus longer and longer systems must be investigated
for a reliable extrapolation. It is worth mentioning, however, that in all cases
the extrapolated curves start with a 
horizontal slope. 
}

\begin{figure}[htb]
\includegraphics[scale=0.65]{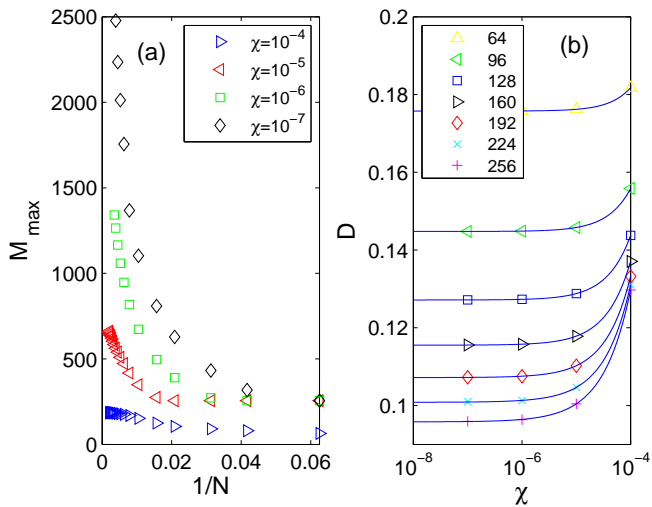}
\caption{(Color online) (a)  
  Maximum number of block states 
  $M_{\rm max}$ selected dynamically for given accuracy thresholds
  $\chi$ plotted as a function of inverse system size $1/N$.  
  (b) Scaling of the dimer order parameter as a function 
  of the quantum information loss $\chi$ for various chain lengths.
}
\label{fig:mmax-n}
\end{figure}

{Since the decrease of the dimer order parameter is the slowest 
at $J_{\bot}=0.38$ we have reinvestigated 
this point using a
better   
extrapolation scheme, i.e., 
the DBSS method.} 
By setting a limit on the quantum information loss, the number 
of block states is selected dynamically during the RG steps of the DMRG. 
In our calculations, the maximum number of block states, $M_{\rm max}$, 
varies in the range $200-2500$ for decreasing $\chi$. 
For $\chi=10^{-4}$, we have taken 
$M_{\rm min}=64$,   
while we taken  $M_{\rm min}=256$ for all other cases. 
The largest dimension of the superblock Hamiltonian was over 14 million.
The number of block states needed to achieve the predetermined $\chi$ value is 
shown as a function of $1/N$ in Fig.~\ref{fig:mmax-n}(a). 
Obviously  $M_{\rm max}$ diverges for increasing $N$ and decreasing
$\chi$. 
For example, $M \simeq 2500$ 
states must be kept for a ladder with $N=256$ rungs in order to obtain
an accuracy of
$\chi=10^{-7}$.

\begin{figure}[htb]
\includegraphics[scale=0.65]{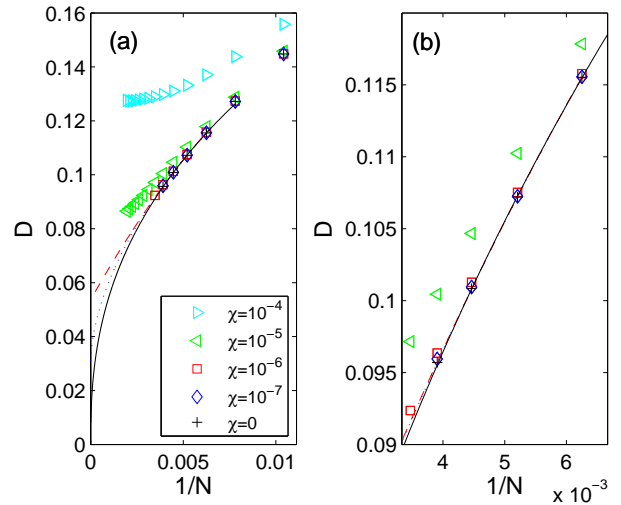}
\caption{(Color online) (a) Similar to Fig.\ \ref{fig:d-m}(b)
  but with data calculated with the indicated fixed values of $\chi$
  and extrapolated to $\chi=0$ using an exponential form as shown in
  \ref{fig:mmax-n}(b). 
The solid line is a fit to the $D(N, \chi\rightarrow 0)$ data 
using Eq.~\eqref{eq:scale_D}. The dotted line is a similar fit without the 
exponential factor, while the dashed line is a fit to a second-order polynomial 
in $1/N$.
(b) Extrapolated data and fits  
for an enlarged 
region.} 
\label{fig:d-n-chi}
\end{figure}

The dimer order parameter calculated for different accuracy levels is
plotted as a function of $\chi$ in Fig.\ref{fig:mmax-n}(b) on a
semi-logarithmic scale, while 
these data and the extrapolated values $D(\chi\rightarrow 0, N)$ 
are plotted as a function of $1/N$ in Fig.~\ref{fig:d-n-chi}(a). 
Extrapolation to $N \rightarrow \infty$ was  again performed using
Eq.~(\ref{eq:scale_D}), yielding  
$D(\chi\rightarrow 0, N\rightarrow\infty) = 0.003(1)$, while the fit 
without the exponential term yielded
$D(\chi\rightarrow 0, N\rightarrow\infty) = 0.021(5)$.  
A portion  
of Fig.~\ref{fig:d-n-chi}(a) is displayed in an enlarged scale in 
Fig.~\ref{fig:d-n-chi}(b). 
It can be clearly seen that
a small error, in this case of the order of $10^{-4}$,
in $D(\chi\rightarrow 0,N)$ can lead to a significant shift of the
extrapolated value.
We have again determined an upper bound through  
a fit to a second-order polynomial
in $1/N$, obtaining $D_{\rm ub}(\chi\rightarrow 0,
N\rightarrow\infty) = 0.054(2)$. 
 
\begin{figure}[htb]
\includegraphics[scale=0.65]{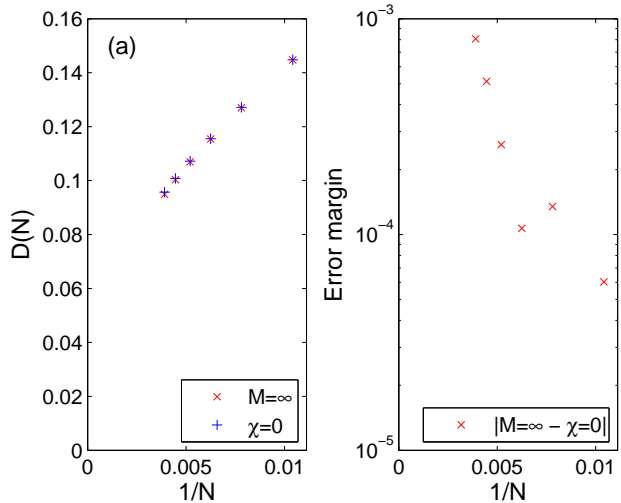}
\caption{(Color online) (a) Similar to Fig.\ \ref{fig:d-m} (b), 
  but with the 
  $M\rightarrow\infty$ and $\chi\rightarrow 0$ extrapolated data
  plotted on the same scale and
  (b) their difference plotted on a logarithmic scale as a function of
  $1/N$.}  
\label{fig:m-chi}
\end{figure}

To compare the results obtained  
from the $M \rightarrow \infty$ extrapolation with those of the
$\chi \rightarrow 0$ extrapolation, we display both
$D(M\rightarrow\infty, N)$ and 
$D(\chi\rightarrow 0, N)$ on the same axes  
in Fig.~\ref{fig:m-chi}(a). 
On the  scale used, the results of the two procedures coincide.
The difference is of order $10^{-4}$, as can be seen on the logarithmic scale
used in Fig.\ \ref{fig:m-chi}  
(b). 
The extrapolated data  
points show no upward curvature  
as a function of $1/N$.
Thus, we find that 
there is no 
evidence for a finite dimer order parameter to the resolution of the
calculations performed here.

\section{Conclusion}

In summary, we have carried out  
density-matrix renormalization-group calculations on the $s=1/2$
cross-coupled spin ladder, in which, in addition to antiferromagnetic
nearest-neighbor couplings along the legs and across the rungs of the ladder, 
an antiferromagnetic coupling between spins diagonally across the
plaquettes is present.
Our aim was to search
for indications for the gapped columnar dimer phase found  
by Starykh and Balents,\cite{oleg} in view of the fact that  
earlier numerical work \cite{liu,hikihara} 
only provided hints that this phase might exist.
Since it is  
known that the dimerized phase,
if it exists, appears in a small parameter regime
due to the marginally relevant character of the current-current
coupling between the legs, 
we have performed   
the numerical calculations using the DMRG 
{ in a narrow region in the parameter space, namely for  
$0.36\leq J_{\bot}\leq 0.4, J_{\times}=0.2$}
and have analyzed the accuracy of the
calculation carefully. 
Whereas the longest ladder studied in Ref.~\onlinecite{hikihara}
had 192 rungs and a maximum of 500 states were kept, we have carried
out calculations for ladders with $N=512$ rungs keeping $M=512$ states
and with $N=256$ rungs keeping $M=1024$ states. 
When the dimer order parameter is plotted as a function of $N$ (see
Fig.~\ref{fig:d-n}), the curves seem to tend to a finite value for
large systems; the slope tends to  
zero. 
This was interpreted in Ref.~\onlinecite{hikihara} 
as an indication of the existence of  
the dimerized phase. 
However, when more and more states are kept, 
this tendency becomes less and less pronounced. 
When  
the dimer order parameter calculated 
for different $M$ values is extrapolated to $M \rightarrow \infty$,
$D$ curves downwards as a function of $1/N$, as seen in
Fig.~\ref{fig:d-m}(b).
{In fact, finite-size scaling 
of the extrapolated $D(M\rightarrow \infty,N)$ data in the whole region 
$0.36\leq J_{\bot}\leq 0.4, J_{\times}=0.2$
using Eq.~\eqref{eq:scale_D} yielded a dimer order parameter of order $10^{-3}$ 
or smaller.  
}

{In addition, 
at a particular
point of the parameter space, $J_{\bot}=0.38,J_{\times}=0.2$, at which the
phase is most likely to occur
}
we have also performed calculations using the DBSS method
taking a number of different 
fixed values for the quantum information loss
$\chi$ for ladders of up to $N=256$ rungs keeping up to  
2500 block states and performing an extrapolation to $\chi=0$. 
One important message of this calculation is that if longer ladders  
are to be considered, the number of states kept 
in the DMRG procedure must be taken to be up to approximately 
10000 to obtain sufficient accuracy.   
This indicates the fundamental limitation  
of the DMRG method for 
the cross-coupled spin ladder.
 
We have found that the result obtained for the dimer order parameter
in the $\chi =0$ limit coincides with the one 
obtained from the $M \rightarrow \infty$ limit. 
The extrapolated data do not show upward curvature as a function of $1/N$, 
in contrast to the results obtained on shorter chains with a smaller number of 
block states kept. 
The limiting value for $N \rightarrow \infty$ depends
strongly on the functional form used for the extrapolation. 
According to our best estimate, the maximum possible value of the
dimer order parameter is of the order of the error of the calculation. 
While it cannot be completely ruled out that the curvature of $D(N)$ 
as a function of $1/N$ changes at much larger system sizes 
if there is  
a very small dimerization gap and a very large correlation length,
there is no indication of such behavior in the present calculations. 
Therefore, we find that whether or not the 
columnar dimer phase is present anywhere 
in the phase diagram of the 
cross-coupled antiferromagnetic ladder remains an open question.

\acknowledgments{
This work was supported in part by the Hungarian Research Fund (OTKA) through 
Grant Nos.~K68340 and K73455. \"O. L. acknowledges support from
the Alexander von Humboldt foundation. The authors acknowledge
computational support from Dynaflex Ltd under Grant No. IgB-32. 
}

\end{document}